\documentclass[aps,pre,showpacs,twocolumn,floats,superscriptaddress]{revtex4}
\usepackage{graphicx}
\usepackage{amsmath}
\usepackage{bm}
\usepackage{color}
\begin{document}
\title{Chaos, Coherence and the Double-Slit Experiment}
\author{Philippe Jacquod}
\affiliation{D\'epartement de Physique Th\'eorique,
Universit\'e de Gen\`eve, CH-1211 Gen\`eve 4, Switzerland}
\affiliation{
Department of Physics, University of Arizona, 1118 E. Fourth Street, 
Tucson, AZ 85721}
\date{September 6, 2005}
\begin{abstract}
We investigate the influence that classical
dynamics has on interference patterns in coherence experiments.
We calculate the time-integrated probability current through an 
absorbing screen and the 
conductance through a doubly connected ballistic cavity, 
both in an Aharonov-Bohm geometry with forward scattering only. 
We show how interference fringes in the 
probability current generically
disappear in the case of a chaotic system with small openings, and how they 
may persist in the case of an integrable cavity. Simultaneously, the 
typical, sample dependent amplitude 
of the flux-sensitive part $g(\phi)$ of the conductance survives
in all cases, and becomes universal in the case of a chaotic cavity. 
In presence of dephasing by fluctuations of the electric potential in one
arm of the Aharonov-Bohm loop, we find an exponential damping of the
flux-dependent part of the conductance,
$g(\phi) \propto \exp[-\tau_{\rm L}/\tau_\varphi]$, 
in term of the traversal time $\tau_{\rm L}$ through the arm
and the dephasing time $\tau_\varphi$. This extends
previous works on dephasing in ballistic systems to the case of 
many conducting channels. 
\end{abstract}
\maketitle

\section{Introduction}
\label{section_intro}
 
Ever since the inception of quantum theory, questions have been raised
related to its connection to classical physics \cite{wheeler83}. From a
dynamical point of view, it is generally accepted that the Liouville and 
Schr\"odinger equations deliver the same time-evolution for 
short enough times, $t < t_{\rm E}$. In both chaotic and integrable
dynamical systems, $t_{\rm E}$ goes
to infinity in the semiclassical limit of large quantum numbers. 
In chaotic systems, however, the quantum breaktime
$t_{\rm E}=\lambda^{-1} |\ln \hbar_{\rm eff}|$
does so only logarithmically slowly in the effective Planck's constant
$\hbar_{\rm eff}$ ($\lambda$ is the system's Lyapunov
exponent) \cite{berman78}. For $t>t_{\rm E}$, the standard view is that 
external sources of decoherence have to be invoked in order to 
reestablish the correspondence between quantum and classical
mechanics \cite{aak,joos,caldeira85,buttiker86a}.

\begin{figure}
\centerline{\hbox{\includegraphics[width=0.95\columnwidth]{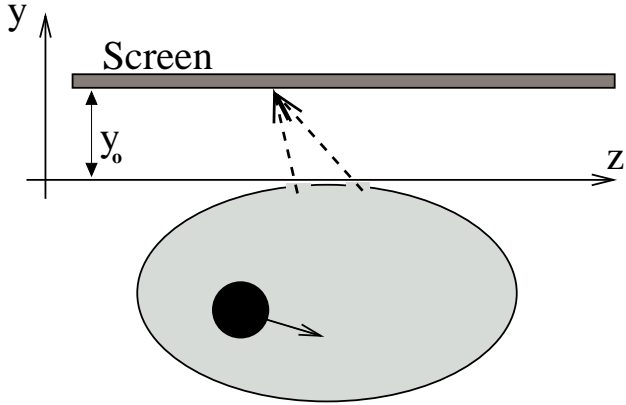}}}
\caption{\label{fig:casati} 
(a) Double-slit set-up of Ref.~\cite{casati04}. A cavity is pierced
by two slits. A wavepacket of well resolved initial momentum
is prepared inside the cavity and leaks out
little by little. The current through
a screen is measured and integrated over time.}
\end{figure}

Arguing that the necessity of external degrees of freedom
for the quantum to classical transition remains unclear
(see for instance Ref.\cite{casati95}),
Casati and Prosen recently performed
a numerical double-slit experiment \cite{casati04}. The set-up
they considered is sketched in Fig.~\ref{fig:casati}. 
One pierces two openings of width $W$ in an otherwise closed cavity. 
Inside the cavity, a particle of mass $m \equiv 1$ is 
prepared in an initial wavepacket of minimal spread in momentum. 
The system is considered to be semiclassical, i.e. the ratio
of the linear system size and the particle's de Broglie wavelength
is big $L/\nu = k L/2 \pi \gg 1$. 
As time goes by, the particle leaks out of the cavity with an average
decay time $\tau_d \propto L^2/(W v) \gg \tau_{\rm f} $
much larger than the time of flight $\tau_{\rm f}\equiv L/v$ across the
cavity, $v$ being the particle's velocity.
That is, the particle bounces many times between the cavity's boundary before
exiting. One then records the integrated probability current $I(x)$ through
the screen, (from now on we set $\hbar \equiv 1$)
\begin{eqnarray}\label{current}
I(z)&=& \int_0^\infty {\rm d}t \; {\rm Im}[ \psi^*(z,y;t) \; \partial_y
\psi(z,y;t) ]_{y=y_{\rm o}}.
\end{eqnarray}
Two different situations were considered, where the cavity was
either integrable (an isosceles right triangle) or chaotic
(where the hypotenuse was replaced by a
circular arc). In the integrable case, numerics showed that $I(x)$ exhibits 
the expected interference fringes. 
Those fringes were however absent in the chaotic case where
$I(x)$ takes on its classical, structureless shape. These results prompted 
Casati and Prosen 
to draw two conclusions: (i) the double-slit set-up provides for 
a ``vivid and fundamental illustration of the manifestation of classical chaos
in quantum mechanics'', and (ii) dynamical chaos alone (i.e. without any
external source of noise, or any coupling to an external bath 
or environment) can produce sufficient
randomization of quantum-mechanical phases resulting in a
quantum to classical transition in the semiclassical limit. 
The reasoning path leading to conclusion (ii) is qualitatively the following.
Due to the long lifetime of the particle
inside the cavity, the wavepacket
must hit the cavity walls many times before exiting. Semiclassically, 
the wavepacket follows many classical trajectories exiting at
different times, and thus accumulating different action phases.
In the regular case, because the particle's initial momentum is
well defined, the action phases accumulated along all those trajectories
are correlated. In the chaotic case
however, the initial momentum uncertainty grows exponentially with time
and the classical trajectories have a broad, continuous
distribution of duration.
Hence they acquire a random distribution of action phases. Based on this
observation, Casati and Prosen concluded that 
this phase randomization prevents interference fringes to 
occur, in agreement with their numerical calculation. It is important
to realize at this point that at any given point and time, the phase
of the wavefunction is uniquely 
defined, and can in principle be deterministically
obtained from the initial condition. 

\begin{figure}
\centerline{\hbox{\includegraphics[width=0.65\columnwidth]{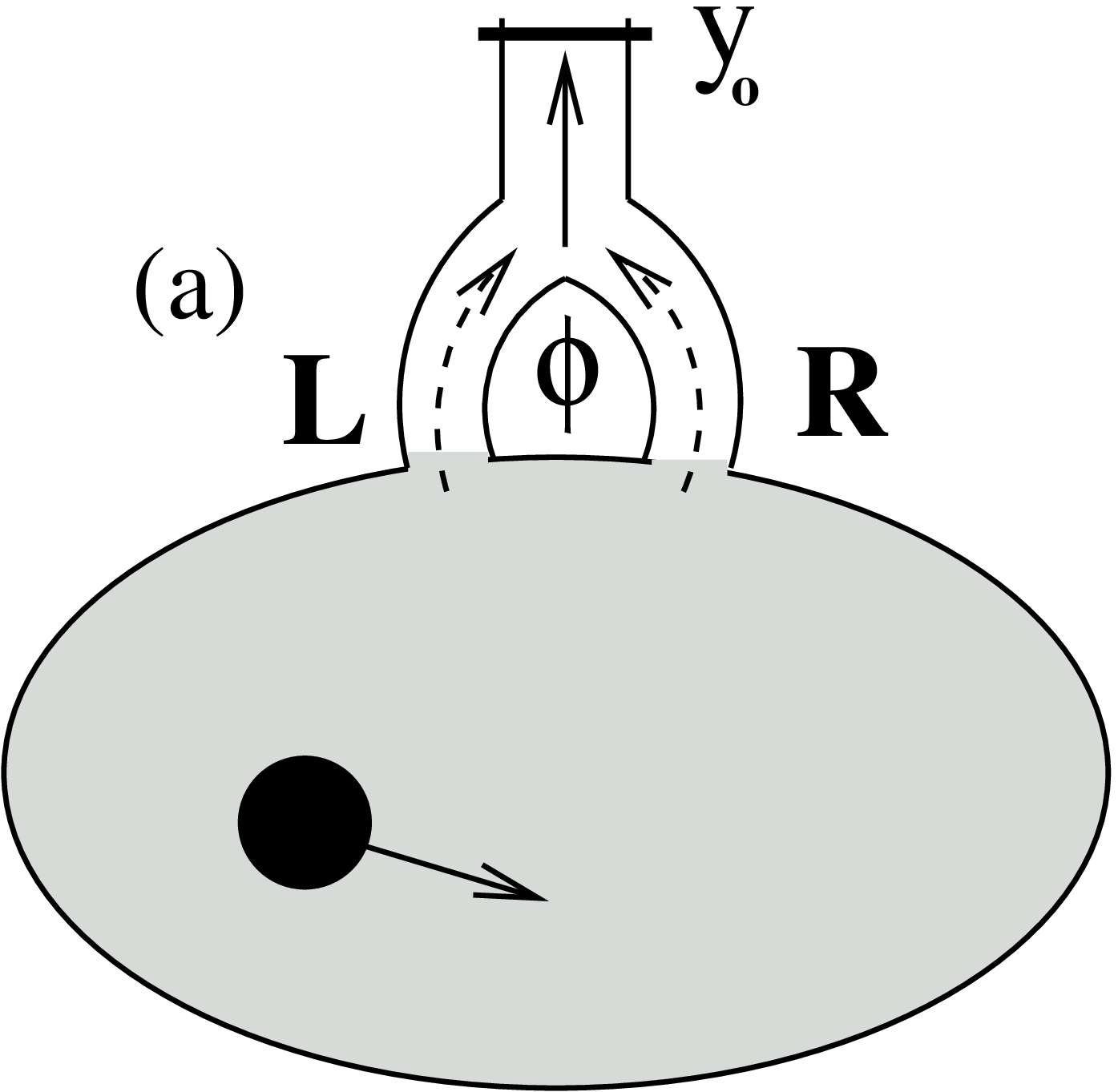}}}
\vskip 0.25truein
\centerline{\hbox{\includegraphics[width=0.65\columnwidth]{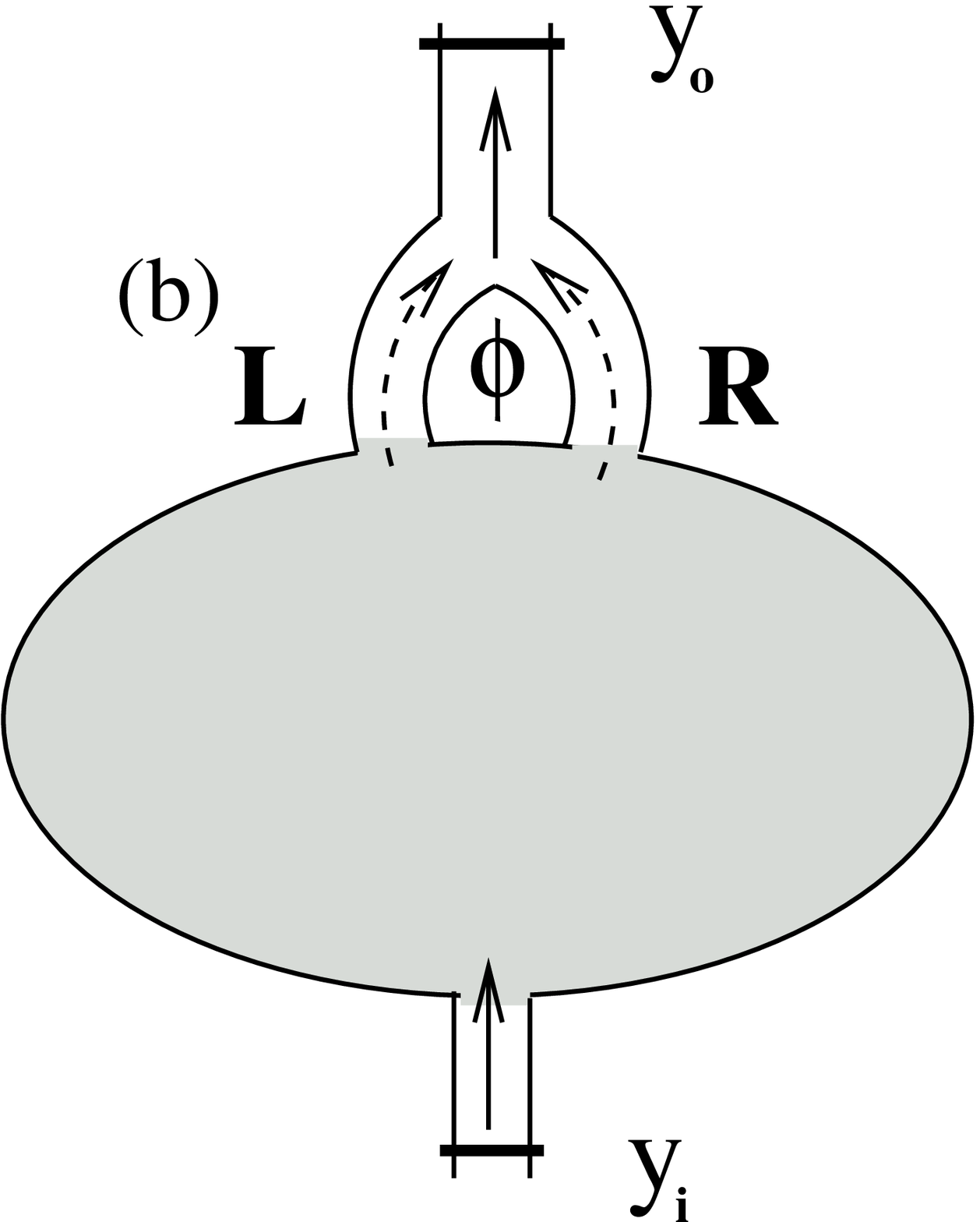}}}
\caption{\label{fig:setups} 
(a) Double-slit set-up similar to that of 
Ref.~\cite{casati04} (see Fig.~\ref{fig:casati}). A cavity is pierced
by two slits which are connected via an Aharonov-Bohm geometry
to a single lead. A wavepacket is prepared inside the cavity and leaks out
little by little. The current through
a cross-section ${\cal C}$ (located at $y=y_{\rm o}$) of the outgoing lead
is measured and integrated over time.
(b) Transport set-up. The same cavity as in (a) is connected
to an additional current-injection lead. The 
current through ${\cal C}$ is measured. 
Taking its ratio with the applied voltage gives the conductance.}
\end{figure}

There is no controversy related to conclusion (i). Conclusion (ii)
however, not only challenges the standard view according
to which long-time quantum-classical correspondence requires coupling
to external degrees of freedom, but has to be reconciled with well established
mesoscopic physics results \cite{textbooks}. It is indeed well known
that both transport \cite{aronov87,webb85,yacoby95} 
and thermodynamical \cite{cheung89,vonoppen91,levy90}
properties of multiply connected mesoscopic samples threaded by 
a magnetic flux display coherent flux-periodic oscillations of a purely
quantal origin. It is doubtful that all experimentally investigated
systems are integrable.
From a theoretical point of view, 
such oscillations have been moreover
predicted for disordered, diffusive samples with 
point-like impurities which are arguably as good ``phase-randomizers''
(in the sense given above) as deterministic chaos.
The even flux-harmonics (those having a period in the applied
flux $\phi$ of $\phi_0/2 n$, with
$n$ a positive integer and $\phi_0=h/e$ the flux quantum)
of these oscillations 
even survive disorder averaging \cite{aronov87,vonoppen91}, and in the case
of transport experiments ``\`a la Sharvin and Sharvin'', 
the amplitude of the Aharonov-Bohm oscillations
of the conductance are mostly insensitive to the amount of disorder
\cite{aronov87}. 
Clearly, conductance is insensitive to the ``dynamical decoherence'' scenario 
of conclusion (ii). The purpose of this
article is to reconcile the numerical experiment of Ref.~\cite{casati04}
with the established 
theoretical and experimental wisdom of mesoscopic physics, as well
as to investigate dephasing in ballistic mesoscopic systems.

We will present a
comparative semiclassical 
calculation of the outgoing probability current in the Aharonov-Bohm 
two-slit set-up of
Fig.~\ref{fig:setups}(a) (similar to the set-up of Ref.~\cite{casati04},
see Fig.~\ref{fig:casati}) and of the conductance in the set-up of
Fig.~\ref{fig:setups}(b). In both cases, a cavity is connected to
two intermediate left (L) and right (R) leads carrying $N_{\rm R,L} \gg 1$ 
transport channels. These intermediate leads eventually merge
and the loop they form is threaded by a magnetic flux $\phi$. In the transport
set-up of Fig.~\ref{fig:setups}(b), 
the cavity is in addition connected to a current-injecting
lead carrying $N_{\rm B}$ transport channels. 
In the two instances, we consider ideally connected,
i.e. non reflecting leads, and will restrict ourselves to the 
situation where the number $N_{\rm T}$ of outgoing channels
obeys $N_{\rm T} \gtrsim N_{\rm L} + N_{\rm R}$. One can then neglect
processes where particles circulate several times around the Aharonov-Bohm
loop. As but one consequence, our semiclassical treatment is fully
unitary, but 
flux-dependent weak localization corrections are absent.
Such corrections have been considered in a different ballistic 
set-up in Ref.~\cite{see03}. The set-up of Fig.~\ref{fig:setups}(b)
in the diffusive regime has been considered in Ref.~\cite{mirlin04}.
A nonunitary semiclassical treatment of the set-up of Fig.~\ref{fig:setups}(b)
considering backscattering due to pairs of time-reversed paths has been
presented in Ref.~\cite{kawabata96}.

Our conclusion is that, while there is nothing
wrong with most of the reasonings and the numerical results of
Ref.~\cite{casati04}, decoherence cannot be claimed to occur 
when one observable
does not display interference patterns, but when this is the case
for all possible observables. 
The conductance experiment of 
Fig.~\ref{fig:setups}(b) will be shown to exhibit sample-dependent
Aharonov-Bohm oscillations in both cases of an integrable and a chaotic
cavity. We will see how these oscillations disappear as dephasing 
is introduced. Our results support the standard wisdom according to which
the quantum to classical crossover requires a coupling to external degrees
of freedom.

The paper is organized as follows. In Section \ref{sec:two-slit} we 
present a semiclassical calculation for an Aharonov-Bohm set-up
similar to the two-slit experiment considered in Ref.~\cite{casati04}.
This calculation is extended to the calculation of the conductance
in an Aharonov-Bohm transport set-up in Section \ref{sec:cond}. In Section 
\ref{sec:dephi} we introduce dephasing by means of a fluctuating
electric potential in one arm of the Aharonov-Bohm loop, and
investigate the associated disappearance of flux-dependent 
interference fringes. In the final Section
\ref{sec:conclusion} we will summarize our findings and discuss future
directions and open questions.

\section{Two-Slit Set-up}\label{sec:two-slit}

We first consider the Aharonov-Bohm two-slit 
set-up of Fig.~\ref{fig:setups}(a),
where an initial wavepacket is prepared inside the cavity. The latter
is connected to two outgoing leads carrying many transverse channels.
The leads eventually merge, forming a loop threaded by a magnetic flux $\phi$.
Once one integrates over a cross-section of the outgoing lead,
the situation is fully similar to Ref.~\cite{casati04}, with $\phi$
playing the role of the coordinate $z$ along the screen 
(see Fig.~\ref{fig:casati}).
We consider an initial Gaussian wavepacket $\psi_0({\bf r}_1) = 
(\pi \nu^2)^{-d/4} \exp[i {\bf p}_0 \cdot ({\bf r}_1-{\bf r}_0)-
|{\bf r}_1-{\bf r}_0|^2/2 \nu^2]$, and approximate its
time-evolution semiclassically by ($H=v^2/2$; remember that we set 
$m\equiv 1$)
\begin{subequations}
\begin{eqnarray}\label{propwp1}
\langle {\bf r}|
\exp(-i H t) |\psi_0\rangle  =  \int d{\bf r}_1
\sum_s K_s({\bf r},{\bf r}_1;t) \psi_0({\bf r}_1),  \\
\label{propwp2}
K_s^{H}({\bf r},{\bf r}_1;t)  =  C_s^{1/2}
\exp[i S_s({\bf r},{\bf r}_1;t)-i \pi \mu_s/2].
\end{eqnarray}\label{propwp}
\end{subequations}
\noindent Compared to Ref.~\cite{casati04}, the Heisenberg uncertainty
is evenly distributed between momentum and spatial coordinates in our
choice of an initial state. This should not matter in a chaotic
cavity, but may affect the outcome of the experiment in a regular cavity.
The semiclassical propagator (\ref{propwp2}) is expressed
as a sum over classical trajectories (labeled $s$)
connecting ${\bf r}$ and ${\bf r}_1$ in the time $t$.
For each $s$, the partial propagator contains
the action integral $S_s^H({\bf r},{\bf r}_1;t)$ along $s$,
a Maslov index $\mu_s$, and
the determinant $C_s$ of the stability matrix.
Because of the cavity openings, if ${\bf r}$ 
in Eqs.~(\ref{propwp}) is inside the cavity, then the sum runs
only over those classical trajectories that have not yet
escaped at time $t$, whereas if ${\bf r}$ lies somewhere in a
lead, it runs over the trajectories
that went exactly once through either of the openings to reach ${\bf r}$. 
Here, we are concerned with this latter case, putting
${\bf r} = (x,y=y_{\rm o})$ at the horizontal position $x$ on 
a cross-section ${\cal C}$ of the outgoing lead defined 
by $y=y_{\rm o}$ [see Fig.~\ref{fig:setups}(a)]. Later on, we will 
integrate over $x$.

The semiclassical expression for the time-integrated probability 
current (\ref{current}) is given by
\begin{eqnarray}\label{sclcurrent}
I(x,\phi)&=& \frac{v}{(\pi \nu^2)^{d/2}} \int_0^\infty {\rm d}t 
\int {\rm d} {\bf r}_1 \int {\rm d} {\bf r}_2 \nonumber \\
& \times & \sum_{s(t),s'(t)} \; [C_s C_{s'}]^{1/2} \cos \theta_s
\nonumber \\
&\times &\exp\Big[i \{S_s(x,y_{\rm o};{\bf r}_1;\phi,t)-
S_{s'}(x,y_{\rm o};{\bf r}_2;\phi,t)\}\Big] 
\nonumber \\
& \times & \exp\Big[i \pi (\mu_{s'}-\mu_s)/2\Big] 
\exp\Big[i {\bf p}_0 \cdot ({\bf r}_1-{\bf r}_2)\Big] 
\nonumber \\
& \times & \exp\Big[-(|{\bf r}_1-{\bf r}_0|^2+|{\bf r}_2-{\bf r}_0|^2)/2 
\nu^2\Big],
\end{eqnarray}
where we used $\partial_y S_s = v_{y,s} = v \cos \theta_s$, 
with $v_{y,s}$ the velocity in $y$-direction and thus 
$\theta_s$ the angle of incidence, as the path $s$ crosses ${\cal C}$ at
time $t$. 

The first step in the calculation of $I(x,\phi)$ is to linearize
$S_s(x,y_{\rm o};{\bf r}_1;\phi,t) \simeq S_s(x,y_{\rm o};{\bf r}_0;\phi,t)-
{\bf p}_s \cdot ({\bf r}_1-{\bf r}_0)$, with 
${\bf p}_s$ the initial momentum on path
$s$. This is justified by our choice of a narrow initial wavepacket.
One it then left with Gaussian integrals over ${\bf r}_{1,2}$.
Enforcing a stationary phase condition,
the dominant, classical contributions to $I(x,\phi)$ are 
identified as those with $s=s'$. Under our assumption of
a final number of transport channels $N_{\rm T} \gtrsim N_{\rm L} +
N_{\rm R}$ roughly equal or 
somehow larger than the sum of transport channels in the
intermediate leads forming the Aharonov-Bohm loop, single trajectories
do not enclose any flux. Diagonal contributions with $s=s'$ 
are thus flux independent. Writing $I(x,\phi) = I_0(x)+I_\phi(x)$, one
has 
\begin{eqnarray}\label{sclcurrentdiag}
\langle I_0(x) \rangle &=& v \; (4 \pi \nu^2)^{d/2} \\
&\times &  \int_0^\infty {\rm d}t 
\sum_{s(t)} \; C_s \cos \theta_s
\exp[-\nu^2 |{\bf p}_0-{\bf p}_s|^2]. \nonumber
\end{eqnarray}
The stationary phase leading to the diagonal approximation $s=s'$
is justified once one averages
over an interval of energy $\delta E$ which is 
classically small (i.e. which does not modify the trajectories)
but quantum-mechanically large (i.e. such that $\delta E \cdot
\tau_d \gg 1$).
This is indicated by
brackets in Eq.~(\ref{sclcurrentdiag}).
The average value $\langle I_0(x) \rangle$ is calculated under
the assumption that the cavity is ergodic, in particular that 
the wavefunction will eventually leak out of it completely. 
That is, the time-integrated
current through ${\cal C}$ must be equal to one and one has
\begin{eqnarray}
\frac{2 v}{\pi} \;
\int_0^W {\rm d} x \int_0^\infty {\rm d}t \; |\psi(x,y_{\rm o};t)|^2 & = 1,
\end{eqnarray}
where a factor $2/\pi$ originated from averaging the incidence angle on
${\cal C}$ in the interval $[-\pi/2,\pi/2]$.
This provides the semiclassical sum rule
\begin{eqnarray}\label{sumrule}
v \; (4 \pi \nu^2)^{d/2}
\int_0^W {\rm d} x \int_0^\infty {\rm d}t && \nonumber \\
\times \sum_{s(t)} \; C_s \exp[-\nu^2 &|{\bf p}_0-{\bf p}_s|^2] =\pi/2.&
\end{eqnarray}
The classical time-integrated current through ${\cal C}$ is then obtained as
\begin{eqnarray}
\int_0^W {\rm d}x \langle I_0(x) 
 \rangle &=& 1.
\end{eqnarray}
In the limit of a wide
outgoing lead, $W \gg \nu$, 
the probability current is ergodically distributed
over ${\cal C}$ so that the average current per unit length is given by
$\langle I_0(x)  \rangle \approx W^{-1}$.

After this warm-up calculation we turn our attention to 
the flux-dependent part $I_{\phi}(x)$. It correspond to 
pairs of paths
$s$ and $s'$ in Eq.~(\ref{sclcurrent}) exiting through different
arms of the AB ring, and evidently they are not included
in the diagonal approximation $s=s'$. Furthermore, no stationary
phase approximation can be systematically enforced to identify
them, which reflects the fact that they vanish on average. That is
to say $\langle I_\phi(x) \rangle = 0$, once it is averaged 
over different initial conditions, a sufficiently large
energy interval or an ensemble of different cavities. This is but one
consequence of our choice of forward scattering processes only
at the merging point of the intermediate leads.

To investigate the behavior of $I_\phi(x)$ for a given cavity and/or
initial wavepacket preparation, we proceed to calculate 
$\langle I^2_\phi (x) \rangle$, the squareroot of which 
gives the value of the flux-dependent part of $I(x)$ for a typical
experimental realization. Our approach is similar in spirit 
to the one followed in Ref.~\cite{cheung89} in the context of persistent
currents. A similar sum rule as (\ref{sumrule}) is helpful in 
computing $\langle I^2_\phi (x) \rangle$, and with little extra work
we will see that $\langle I^2_{\phi}(x) \rangle
\propto (1-\exp[-\tau_{\rm erg}/\tau_d])^2 (k L)^{-1}
\approx (\tau_{\rm erg}/\tau_d)^2 (k L)^{-1}$,
where $\tau_{\rm erg}$ is the ergodic time. In a chaotic cavity, it
is generally given by few times the time of flight across
the cavity, so that $\tau_{\rm erg}/\tau_d \propto W/L$. In the
numerical experiment of Ref.~\cite{casati04}, both the ratio of the width
of the openings to the linear system size and the inverse semiclassical
parameter $k L$ are much smaller
than one, inducing the disappearance of the interference fringes.

\begin{widetext}
Noting that $S_s(x,y_{\rm o};{\bf r}_0;\phi,t) = S_s(x,y_{\rm o};{\bf r}_0;t) 
\pm \pi \phi/\phi_0$, where the $''+''$ and $''-''$ signs correspond 
to trajectories going through the right and 
the left intermediate lead respectively,
linearizing in ${\bf r}_{1,2}-{\bf r}_0$ and performing the resulting Gaussian
integrals over ${\bf r}_{1,2}$ as above, one has
\begin{eqnarray}\label{sclcurrentphi}
I_{\phi}(x) I_{\phi}(x') &=& 4 v^2 \;
(4 \pi \nu^2)^{d}
\cos^2 [2 \pi \phi/\phi_0] \;\int_0^\infty {\rm d}t_1  \int_0^\infty {\rm d}t_2
\; \sum_{s_1,s_3 \in L} 
 \sum_{s_2,s_4 \in R} \; \left(\; \prod_{i=1}^4 C_i \;\right)^{1/2} \cos \theta_1 \;
\cos \theta_4
\nonumber \\
&\times &\exp\Big[i \big\{S_1(x,y_{\rm o};{\bf r}_0;t_1)-S_2(x,y_{\rm o};{\bf r}_0;t_1)
-S_3(x',y_{\rm o};{\bf r}_0;t_2)+S_4(x',y_{\rm o};{\bf r}_0;t_2)\big\}
\Big] 
\nonumber \\
& \times & \exp\Big[-\nu^2\big\{|{\bf p}_0-{\bf p}_1|^2+
|{\bf p}_0-{\bf p}_2|^2+|{\bf p}_0-{\bf p}_3|^2+|{\bf p}_0-{\bf p}_4|^2 
\big\}/2
\Big],
\end{eqnarray}
where we shortened the notation, i.e. $\theta_i=\theta_{s_i}$,
$S_i=S_{s_i}$ and so forth. It is important to keep in mind that $s_1$
and $s_2$ exit the cavity after a time $t_1$, while the time of escape
is $t_2$ for the other two trajectories
$s_3$ and $s_4$. Because trajectories exit via two different
arms of the AB loop, the only stationary phase condition that
can be satisfied is to set $s_1=s_3$ and $s_2=s_4$, which then
requires to set $x=x'$ with accuracy $\nu$, and 
$t_1 \approx t_2$, with an accuracy given by the time
$\tau^* \simeq \hbar/E$ necessary for the classical 
ballistic flow 
at energy $E$ to accumulate an action $\hbar$.
We substitute $\int {\rm d}t_1 \int {\rm d}t_2  \rightarrow
\tau^* \int {\rm d}t_1 $
to get
\begin{equation}\label{iphi}
\langle I_{\phi}(x) I_{\phi}(x') \rangle \simeq
\delta_\nu(x-x') \; 4 v^2 \tau^* \;
(4 \pi \nu^2)^{d}
\cos^2 [2 \pi \phi/\phi_0] \int_0^\infty {\rm d}t_1 
\; \sum_{s_1\in L} 
\sum_{s_2\in R} \; \prod_{i=1}^2 \Big\{ C_i \cos \theta_i \;
\exp\big[-\nu^2|{\bf p}_0-{\bf p}_1|^2\big] \Big\},
\end{equation}
where $\delta_\nu(x-x')$ enforces the condition
$x=x'$ with an accuracy ${\cal O}(\nu)$.
Because there is only one time integral but two summations over
classical paths, one cannot use Eq.~(\ref{sumrule})
directly. Assuming that
the system is ergodic, which means in particular 
that for times long enough,
$t \ge \tau_{\rm erg} \approx \tau_{f}$, spatial averages equal time averages,
one writes
\begin{eqnarray}\label{iphi2}
\langle I^2_{\phi}(x) \rangle & \simeq & 4 v^2 \;
(4 \pi \nu^2)^{d}
\cos^2 [2 \pi \phi/\phi_0] \Bigg[ \tau^* \;\int_0^{\tau_{\rm erg}} {\rm d}t_1 
\; \sum_{s_1\in L} 
 \sum_{s_2\in R} \; \prod_{i=1}^2 \Big\{ C_i \cos \theta_i \;
\exp\big[-\nu^2|{\bf p}_0-{\bf p}_1|^2\big] \Big\} \nonumber \\
&+& \;\int_{\tau_{\rm erg}}^\infty
 {\rm d}t_1 
\lim_{T \rightarrow \infty} \frac{\tau^*}{T} \int_{\tau_{\rm erg}}^T
 {\rm d}t_2
\; \sum_{s_1\in L} 
 \sum_{s_2\in R} \; \prod_{i=1}^2 \Big\{ C_i \cos \theta_i \;
\exp\big[-\nu^2|{\bf p}_0-{\bf p}_1|^2\big] \Big\}\Bigg].
\end{eqnarray} 
Here, the second term inside the brackets corresponds to trajectories
$s_1(t_1)$ and $s_2(t_2)$ exiting at different times. 
Its contribution to the integrated
current $\int dx \langle I^2_{\phi}(x) \rangle$
can be calculated using the sum rule
(\ref{sumrule}) and making the assumption that
the current is homogeneously distributed on ${\cal C}$.
We find that it vanishes $\propto
{\rm lim}_{T\rightarrow \infty} \; \tau^*/T$.
The first, pre-ergodic term is highly non-universal
and we cannot calculate it generically. We can however give an estimate
to its amplitude using
\begin{eqnarray}
\;\int_0^{\tau_{\rm erg}} {\rm d}t_1 \; f(t_1) \; g(t_1)
\sim \tau_{\rm erg}^{-1} \;\int_0^{\tau_{\rm erg}} {\rm d}t_1 \;
{\rm d}t_2 \; f(t_1) \; g(t_2)
\sim \tau_{\rm erg}^{-1} \; (1-\exp[-\tau_{\rm erg}/\tau_d])^2  
\;\int_0^\infty {\rm d}t_1 \;
{\rm d}t_2 \; f(t_1) \; g(t_2).
\end{eqnarray}
The first relation results from removing the requirement that both 
trajectories $s_1$ and $s_2$ in Eq.~(\ref{iphi2}) exit at the same time,
and to obtain the second one,
we used the measure of pre-ergodic trajectories in an open
chaotic cavity
$\tilde{\rho}(t \le \tau_{\rm erg}) = \tau_d^{-1}
\int_0^{\tau_{\rm erg}} {\rm d} t \exp[-t/\tau_d]$, where
$\rho(t)= \tau_d^{-1} \exp[-t/\tau_d]$ is the distribution of dwell times 
through a chaotic system \cite{Bauer90}.
Using $\tau^*/\tau_{\rm erg} \simeq (k L)^{-1}$, and assuming again 
an homogeneous distribution of $I(x)$ on ${\cal C}$,
we finally get the typical flux-dependent probability current as
\begin{eqnarray}\label{estimate_iphi}
\langle I^2_{\phi}(x) \rangle^{1/2}& \sim &  
\cos [2 \pi \phi/\phi_0] \; 
(1-\exp[-\tau_{\rm erg}/\tau_d])  \; 
\sqrt{\frac{\tau^*}{\tau_{\rm erg}}} \; \langle I_0(x) \rangle.
\end{eqnarray}
\end{widetext}

We believe that Eq.~(\ref{estimate_iphi}) gives an upper bound for the
typical flux-dependent part of the probability current in the case of a chaotic
cavity. One sees that, compared to $\langle I_0 \rangle$,
$\langle I_\phi^2\rangle^{1/2}$ is suppressed by a prefactor
$(1-\exp[-\tau_{\rm erg}/\tau_d]) (k L)^{-1/2}$. 
In the chaotic configuration of Ref.~\cite{casati04},
the dwell time is approximately several hundreds of times larger
than the ergodic time. Together with $k L = 180$, this 
leads to the suppression of 
the flux oscillations in a given sample by a relative factor of at least
$\propto (\tau_{\rm erg}/\tau_d) (k L)^{-1/2} \le 10^{-3}$ compared
to the average current value. 

While it is always risky to make generic statistical 
statements on regular systems, it is reasonable to expect that
in this case, the pre-ergodic terms in Eq.~(\ref{iphi})
provide for most of the contributions to $\langle I^2_{\phi}(x) \rangle$. 
This is so, since for regular systems,
$\tau_{\rm erg}$ is much larger than in a chaotic system,
and even diverges in most instances, regular systems being usually
not ergodic. Moreover, integrable systems exhibit periodicities
and quasiperiodicities and a persistence of correlations over very large
times. Starting from Eq.~(\ref{sclcurrentphi}), one may thus
pair trajectories either with $\tau^* \gg \hbar/E$, or even completely 
relaxing the restriction $|t_1-t_2| \lesssim \tau^*$. One then gets the 
best case scenario result that
\begin{eqnarray}\label{estimate_iphireg}
\langle I^2_{\phi}(x) \rangle_{\rm reg}^{1/2}& \sim &  
\cos [2 \pi \phi/\phi_0] \; \langle I_0(x) \rangle,
\end{eqnarray}
i.e. the flux-dependent probability
current is of the same magnitude as its classical part $I_0(x)$.
This is also in agreement with Ref.~\cite{casati04}. One should stress however
that the result (\ref{estimate_iphireg}) cannot be expected to hold 
generically. In particular, we believe that the choice made
in  Ref.~\cite{casati04} of an initial state
with narrowest momentum spread is necessary to get
interference fringes satisfying (\ref{estimate_iphireg}).
Presumably the choice of direction of momentum also plays a role.

To summarize this section, we have shown why the interference fringes
disappear for a two-slit experiment out of a chaotic cavity.
The main result of this section, Eq.~(\ref{estimate_iphi}), can be checked 
numerically by increasing the width $W$ of the slits
or varying $kL$, or both, in the numerical experiment 
of Ref.~\cite{casati04}. 
More qualitatively, we argued that in well chosen situations,
the interference fringes have a
magnitude comparable to the classical probability current if the cavity
is regular.\\

\section{Transport Set-up}\label{sec:cond}

We next focus on the transport set-up shown in Fig.~\ref{fig:setups}(b).
We write the conductance as a sum of a classical nd 
a flux-dependent part, $g = g_0 + g(\phi)$. 
We use the semiclassical framework developed in 
Ref.~\cite{Bar93}. We start from the scattering
approach which relates transport properties to the
system's scattering matrix \cite{scatg} 
\begin{eqnarray}\label{blocks}
{\cal S}= \left( \begin{array}{ll}
{\bf r} & {\bf t}' \\
{\bf t} & {\bf r}' 
\end{array}\right).
\end{eqnarray}
For the two terminal geometry we consider,
${\cal S}$ is a 2-block by 2-block matrix, written in terms of transmission 
(${\bf t}$ and ${\bf t}'$)
and reflection (${\bf r}$ and ${\bf r}'$) matrices. 
From ${\cal S}$, the system's conductance is 
given by $g={\rm Tr} ({\bf t}^\dagger {\bf t})$
($g$ is expressed in units of $e^2/h$). 

From Ref.~\cite{Bar93}, the matrix elements $t_{mn}$ of the transmission matrix
${\bf t}$ are written as 
\begin{equation}\label{trmel}
t_{mn} = -\sqrt{\frac{\pi \hbar}{2 W_{\rm B} W_{\rm T}}} \;
\sum_{s} \frac{\; \Phi_s \; \exp[i S_s({\bf r}_{\rm B},{\bf
      r}_{\rm T};E)] \;}{|\cos \theta_{\rm B}^{(m)} 
\cos \theta_{\rm T}^{(n)} M_{21}^s |^{1/2}}.
\end{equation} 
The sum runs over all classical scattering trajectories 
entering the cavity with an angle $\pm \theta_{\rm B}^{(m)}$ at
any point ${\bf r}_{\rm B}=(x,y_{\rm i})$ 
on a cross-section ${\cal C}_{\rm B}$ of
the bottom lead (of geometric width $W_{\rm B}$) and exiting it
with an angle $\pm \theta_{\rm T}^{(n)}$
at any point ${\bf r}_{\rm T}=(x',y_{\rm o})$ on a cross-section 
${\cal C}_{\rm T}$ of the top lead (of geometric width $W_{\rm T}$). 
The channel indices $(m,n)$ specify the entrance and exit angles as
$\sin \theta_{\rm B}^{(m)}= \pi \bar{m}/k_F W_{\rm B}$ and
$\sin \theta_{\rm T}^{(n)}= \pi \bar{n}/k_F W_{\rm T}$, 
$\bar{m}=\pm m$, $\bar{n}=\pm n$, while  
$S_s({\bf r}_{\rm B},{\bf r}_{\rm T};E)$ gives the classical
action accumulated along $s$. Finally
$M_{21}^s=\partial v_\perp/d q_\perp$ is an element of the
monodromy matrix (the $\perp$-direction is normal to the cross-sections), 
and there is a phase factor
$\Phi_s={\rm sgn}(\bar{m}) {\rm sgn}(\bar{n}) \exp[i \pi (\bar{m} 
x_{\rm B}/W_{\rm B} -\bar{n} x_{\rm T}/W_{\rm T}-\mu_s/2+1/4)]$. 

\begin{widetext}
All one needs to calculate the average conductance of a chaotic
cavity is the following sum rule, 
valid in the regime of classical ergodicity \cite{Bar93}
\begin{eqnarray}\label{sumrule2}
\sum_{s(x_{\rm B},x_{\rm T};\theta_{\rm B}^{(m)},\theta_{\rm T}^{(n)})} 
\frac{\delta(\tau-\tau_s)}{| M_{21}^s |} \simeq 
\frac{\cos \theta_{\rm B}^{(m)} \cos \theta_{\rm T}^{(n)}}{\Sigma(E)} \;
\delta x_{\rm B} \delta x_{\rm T} \; \tilde{\rho}(\tau).
\end{eqnarray}
In contrast to Eq.~(\ref{trmel}), the sum in Eq.~(\ref{sumrule2})
is restricted to phase-space trajectories with a well resolved 
position and momentum direction on 
${\cal C}_{\rm B}$ and ${\cal C}_{\rm T}$, up to uncertainties
$\delta x_{\rm B}, \delta x_{\rm T} \simeq \nu$. 
Here, $\Sigma(E)=2 \pi A$ gives the volume of phase space that can be
visited by an ergodic particle of energy $E$ in a cavity of area $A$,
and $\tilde{\rho}(\tau)
=\int_\tau^\infty \rho(t) {\rm d} t= \exp[-\tau/\tau_d]$ gives
the survival probability that a particle remains
inside an open chaotic system for a time longer than, or equal to $\tau$.
The meaning of the sum rule
(\ref{sumrule2}) is that at any time, surviving classical trajectories
have a probability to exit the cavity given by the fraction of phase-space
volume covered by the leads to the total accessible volume of phase-space.

From Eqs.~(\ref{trmel}) and (\ref{sumrule2}), together with
the relation $\tau_d = \pi A/[v (W_{\rm B}+W_{\rm T})]$,
it is straightforward
to calculate the average conductance within the diagonal approximation.
One ends up with the classical conductance
\begin{eqnarray}\label{gcl}
\langle g \rangle & = & \sum_{m,n} \frac{\pi \hbar}{2 W_{\rm B} W_{\rm T}} \;
\sum_{s} \Big|\cos \theta_{\rm B}^{(m)} 
\cos \theta_{\rm T}^{(n)} M_{21}^s \Big|^{-1} 
 =  \frac{N_{\rm B} \cdot N_{\rm T}}{N_{\rm B} + N_{\rm T}},
\end{eqnarray}
\end{widetext}
where we used the relation between lead width and channel number
$N = {\rm Int}[k_{\rm F} W/\pi]$.
As was the case for the probability current, the average
conductance has no flux dependence since diagonally paired
trajectories do not enclose any flux.

Following the procedure we applied to $\langle I^2(\phi) \rangle$,
it is straightforward to calculate the squared typical value 
of the flux-dependent part of the conductance 
$\langle g^2(\phi) \rangle$ using Eqs.~(\ref{trmel}) and (\ref{sumrule2}),
and 
$S_s(x,y_{\rm o};{\bf r}_0;\phi,t) = S_s(x,y_{\rm o};{\bf r}_0;t) 
\pm \pi \phi/\phi_0$. One then has
\begin{eqnarray}\label{gphi}
\langle g^2(\phi) \rangle & = & 
\frac{16 \pi^2 \hbar^2 N_{\rm B} \; N_{\rm T}}{\Sigma^2(E)} 
\left(\int_0^\infty {\rm d}t \tilde{\rho}(t) \right)^2 \cos^2(2 \pi
\phi/\phi_0) \nonumber \\
& = & 
4 \frac{N_{\rm B} \; N_{\rm T}}{(N_{\rm B} + N_{\rm T})^2}
\cos^2(2 \pi \phi/\phi_0).
\end{eqnarray}
Compared to the square of 
Eq.~(\ref{gcl}), one sum over pairs of channel indices disappeared
from Eq.~(\ref{gphi}) because of the stationary phase condition
we enforced on each of the two pairs of orbits going through the left and
right intermediate lead respectively.

Eq.~(\ref{gphi}) is the main result of this section. It shows the
universality of the typical Aharonov-Bohm response of the conductance
in our set-up in the chaotic case. For $N_{\rm B}$ and $N_{\rm T}$ not
too different from each other, $\langle g^2(\phi) \rangle$ is 
independent on $\langle g \rangle$. The survival of interference fringes in the
transport set-up is a direct consequence of the fact that
to extract the conductance, one works in energy representation.
Once one writes the scattering
matrix in time representation, the squared typical conductance is given by an
expression similar to Eq.~(\ref{sclcurrent}), with however two time
integrals. This makes it much easier to extract stationary 
phase conditions, without going through
the ergodicity tricks that were needed to go from
Eq.~(\ref{sclcurrentphi}) to Eq.~(\ref{estimate_iphi}), and explains
the ease of calculation with which (\ref{gphi}) is derived compared to its
probability current counterpart of Eq.~(\ref{estimate_iphi}). 

As was the case in the previous section for the probability current, 
we cannot calculate $\langle g^2(\phi) \rangle$
in the integrable case without relying on assumptions which are not
necessarily well controlled.
In particular, there is, to the best of our knowledge, no sum rule
such as (\ref{sumrule2}) for regular systems. 
As is the case for persistent currents
in ballistic systems however \cite{vonOp93}, one expects 
a significantly increased magnetic response,
well above the chaotic value (\ref{gphi}),
because in a regular system,
the dwell time distribution is not exponential, but power-law
 $\rho(\tau) \propto \tau^{-\beta}$ \cite{Bauer90}.
In the best case scenario, one can expect a response given by the coherent sum
of $N$ responses [$N={\rm min}(N_{\rm B},N_{\rm T})$], leading to
a flux dependence of a similar amplitude as the conductance itself.
Here, further numerical experiments are needed to clarify the situation.

\section{Dephasing by a Fluctuating Potential}\label{sec:dephi}

The results (\ref{estimate_iphi}) and 
(\ref{gphi}) derived above follow from a stationary phase condition.
To satisfy the latter, one relies on the exact pairing of trajectories,
i.e. setting $s=s'$ where applicable, and in this way, all accumulated action 
phases cancel two by two. This is no longer the case in the presence
of an external dephasing source. In this case, 
phase differences inevitably occur
in pairs of contracted trajectories
due to the interaction with the external source
of noise at different times along the trajectory. In this section,
we finally discuss this occurrence and how dephasing destroys
the Aharonov-Bohm interference fringes. 

\begin{figure}
\centerline{\hbox{\includegraphics[width=0.95\columnwidth]{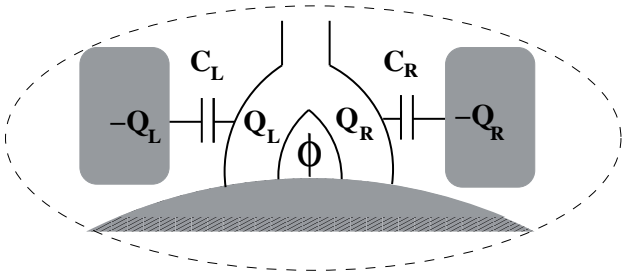}}}
\caption{\label{fig:dephi} Aharonov-Bohm loop capacitively
coupled to external charged gates. 
The system as a whole (inside the
dashed line; this includes the full cavity which we drew only
partially) is electrically neutral, which does not prevent 
charge fluctuations in the arms of the loop to occur, provided they
are compensated by fluctuations in the gates. The fluctuations in 
$Q_{\rm L}$ and $Q_{\rm R}$ induce fluctuations of the internal
electric potential in the corresponding arm.}
\end{figure}

Following Ref.~\cite{see01},
we consider that our system as a whole, including charged gates defining
the cavity and the Aharonov-Bohm ring, is electrically neutral, as sketched
in Fig.~\ref{fig:dephi}. This does
not prevent local charge fluctuations to occur, which in their turn
induce fluctuations of the electric potential felt by the electrons.
This is a specific example of dephasing induced by an external source,
in this case the electric charges on the gates defining
the system, which must fluctuate to ensure that the fluctuations inside
the circuit are compensated to make the whole system electrically neutral.
These fluctuations result in dephasing, and
without loss of generality, we will assume that they affect 
only electrons passing through
one, say the left intermediate lead, during the traversal time $\tau_{\rm L}
=L_{\rm L}/v$ through that lead.

We consider the case of weak coupling, where the trajectories are 
unaffected by the coupling to external degrees of freedom. Dephasing is
introduced in our calculation via the substitution
\begin{equation}
S_s(x,y_{\rm o};{\bf r}_0;t) \rightarrow 
S_s(x,y_{\rm o};{\bf r}_0;t) 
+ \int_0^{\tau_{\rm L}} dt \; \varphi_s(t).
\end{equation}
Here $\varphi_s(t)$ gives the additional
action phase accumulated by an electron traveling on path $s$ 
and interacting with the dephasing source at time $t$.

Using the central limit theorem, 
Eqs.~(\ref{estimate_iphi}) and (\ref{gphi}) have then to be multiplied by 
\begin{equation}\label{exp_dphi}
\exp[-\int_0^{\tau_L} dt_1 \int_0^{\tau_L} dt_2
\langle \varphi_s(t_1) \varphi_s(t_2) \rangle_s /2],
\end{equation} 
where $\langle ... \rangle_s$
denotes an average over the distribution of phases on 
different classical trajectories. 
Further assuming an exponential decay of the phase 
correlator $\langle \varphi_s(t_1) \varphi_s(t_2) \rangle_s
= \langle \varphi_s^2(0) \rangle_s \exp[-|t_1-t_2|/\tau_c]$,
one gets, in the limit $\tau_c \ll \tau_{\rm L}$, an exponential 
suppression of the flux response 
\begin{eqnarray}\label{damping}
\langle g^2(\phi) \rangle & = & 
\frac{N_{\rm B} \; N_{\rm T}}{(N_{\rm B} + N_{\rm T})^2}
\cos^2(2 \pi \phi/\phi_0) e^{-\tau_{\rm L}/\tau_\varphi},
\end{eqnarray}
where $\tau_\varphi^{-1} = 2 \tau_c \langle \varphi_s^2(0) \rangle_s$. 
In the limit of Nyquist noise, a self-consistent calculation
of the phase correlator has been performed in Ref.~\cite{see01}, 
within the one-potential approximation, i.e. assuming that the fluctuations
of the electric potential are spatially homogeneous inside one arm.
A linear temperature dependence of the dephasing rate was obtained, which
in our case translates into
\begin{equation}\label{tdphi}
\tau_\varphi^{-1} = 2 \tau_c \langle \varphi_s^2(0) \rangle_s = 8 \pi
\gamma^2_{\rm L} k_B T/N_{\rm L}.
\end{equation}
Here, $\gamma_{\rm L} \le 1$ 
stands for the ratio between the electrochemical  
and the electrical capacitance of the left arm \cite{see01}. In the
weak coupling limit we are considering, one has $\gamma_{\rm L} \simeq 1$.
Both the exponential damping of the Aharonov-Bohm flux and the
linear temperature dependence of the dephasing rate are
in agreement with the experimental results of
Ref.~\cite{han01} on Aharonov-Bohm conductance oscillations in
few-channel ballistic systems. Our results (\ref{exp_dphi})-(\ref{tdphi})
extend those of Ref.~\cite{see01}
to the many-channel case.

As a side remark, we note that in  
the other limit $\tau_c \gg \tau_{\rm L}$, one gets a Gaussian
suppression of the flux response in the traversal time $\tau_{\rm L}$,
\begin{eqnarray}
\langle g^2(\phi) \rangle & = & 
\frac{N_{\rm B} \; N_{\rm T}}{(N_{\rm B} + N_{\rm T})^2}
\cos^2(2 \pi \phi/\phi_0) e^{-\tau^2_{\rm L}/\tau_\varphi \tau_c},
\end{eqnarray}
with the same dephasing time as above. This Gaussian damping has not
been obtained previously. Indeed, previous works always assumed
$\delta$-correlated phases,  $\langle \varphi_s(t_1) \varphi_s(t_2) \rangle_s
\propto \delta(t_1-t_2)$, meaning $\tau_c/\tau_{\rm L} \rightarrow 0$.

To close this chapter, we remark that the same dephasing behavior will
occur in regular systems as long as the phase correlator decay fast enough.
While in that case, an exponential decay is not at all obvious
from a dynamical point of view, we stress that, 
in the limit of long traversal times $\tau_{\rm L} \gg \tau_c$,
the minimal requirement for an exponential damping as
in Eq.~(\ref{damping}) is a power law decay of the phase correlator
$\langle \varphi(t_1) \varphi(t_2) \rangle
= \langle \varphi(0) \varphi(0) \rangle [\tau_c/(\tau_c+|t_1-t_2|)]^{\alpha}$
with $\alpha > 1$.\\[-2mm]

\section{Conclusion}\label{sec:conclusion}

We have presented a semiclassical calculation of the
flux dependence of the probability current and the conductance in
two distinct Aharonov-Bohm set-ups (see Fig.~2). We have shown
how the interference fringes in the 
probability current disappear in chaotic systems in the
case of cavities with large dwell times, whereas they may persist in the
case of a regular cavity. This is in agreement with and 
sheds light on the numerical results of Ref.~\cite{casati04}.
Simultaneously, we showed how the situation is completely different 
in the transport set-up, where the flux response of the conductance
becomes universal in the chaotic case. This universality is
lost in the case of integrable cavities, where we conjectured
that the flux response may
be of the same order as the conductance itself.

In the transport set-up, we argued
that dephasing from external degrees of freedom is
necessary to wash out the flux-periodic interference fringes in
the conductance.
We introduced dephasing in a similar way as in Refs.~\cite{see03,see01}
and found that flux-dependent interference fringes in the conductance
vanish exponentially, $\exp[-\tau_{\rm L}/\tau_{\varphi}]$. 
Both this exponential damping
and the linear temperature-dependence of the dephasing time
(\ref{tdphi}) are in
agreement with transport experiments on ballistic Aharonov-Bohm 
systems \cite{han01}. 
Our results confirm the standard view that external sources of
decoherence are generally required to induce a complete quantum--classical
correspondence. 

Our semiclassical treatment extends the results
of Refs.~\cite{see03,see01} to the many-channel case. Still,
the dephasing behavior of Eqs.~(\ref{exp_dphi})-(\ref{damping}) relies on
the one-potential approximation giving the linear temperature dependence
of the phase correlator, Eq.~(\ref{tdphi}). Because Ref.~\cite{casati04} 
considered
the other limit of sub-wavelength slits, it is likely that
diffraction effects play a role there that was neglected here.
However, we do not expect diffraction to alter the situation 
qualitatively. 

One of our 
motivations was to reconcile the results of Ref.~\cite{casati04}
with well-known mesoscopic physics theoretical and experimental results.
That is why we deliberately made the hypothesis of forward scattering only, 
that particles entering one of 
the intermediate leads (indicated by $L$ and $R$ in Fig.~2) are transferred
to the outgoing lead with probability one. This is justified
in the case where that latter lead is somehow wider than the
two intermediate leads together, $N_{\rm T} \gtrsim N_{\rm R}+N_{\rm L}$.
It would be interesting to lift that hypothesis, and consider the emergence of 
higher flux harmonics and of flux-dependent
weak localization corrections to the average conductance, 
and the influence that dephasing has on them.
We expect that the
presence of weak-localization corrections would result in the usual Lorentzian 
damping of the amplitude of Aharonov-Bohm
interference fringes in the disorder-averaged conductance (as opposed to the
typical conductance calculated here). 
Further investigations are however
necessary to confirm this.

\section*{Acknowledgments}
This work has been supported by the Swiss National Science Foundation.
It is a pleasure to thank M. B\"uttiker for drawing our attention
to Ref.\cite{casati04} and for several interesting discussions and comments.

\end{document}